\documentclass[12pt]{article}
\usepackage{epsfig}
\usepackage{epsf}
\begin{document}
\pagestyle{plain}
\date{}


\title{Chaotic dynamics in a storage-ring Free Electron Laser}

\author{G. De Ninno$^{1,2,*}$, D. Fanelli$^{3}$, C. Bruni$^{1,2}$, 
M.E. Couprie$^{1,2}$}

\date{}
\maketitle

\begin{center}
\begin{tabular}{ll}

$^1$ &CEA, DSM/DRECAM/SPAM, Cea-Saclay 91 191 Gif sur Yvette, France. \\    
$^2$ & LURE, Bat. 209 D Universit\'e de Paris-Sud 91 405 Orsay Cedex,
France.\\
$^3$ & NADA, KTH, SE-100 44 Stockholm, Sweden.\\
$^*$ & Presently at: Sincrotrone Trieste, 34012 Trieste, Italy.
\end{tabular}
\end{center}

\normalsize

\begin{abstract}
The temporal dynamics of a storage-ring Free Electron Laser is here 
investigated with particular attention to the case in which an 
external modulation is applied to the laser-electron beam detuning.
The system is shown to produce bifurcations, multi-furcations as well 
as chaotic regimes. The peculiarities of this phenomenon with respect 
to the analogous behavior displayed by conventional laser sources are
pointed out. Theoretical results, obtained by means of a 
phenomenological model reproducing the evolution of the main 
statistical parameters of the system, are shown to be in a good 
agreement with experiments carried out on the 
Super-ACO Free Electron Laser.
\end{abstract}


\section{Introduction}

\begin{figure}[ht!]
\centering
\epsfig{file=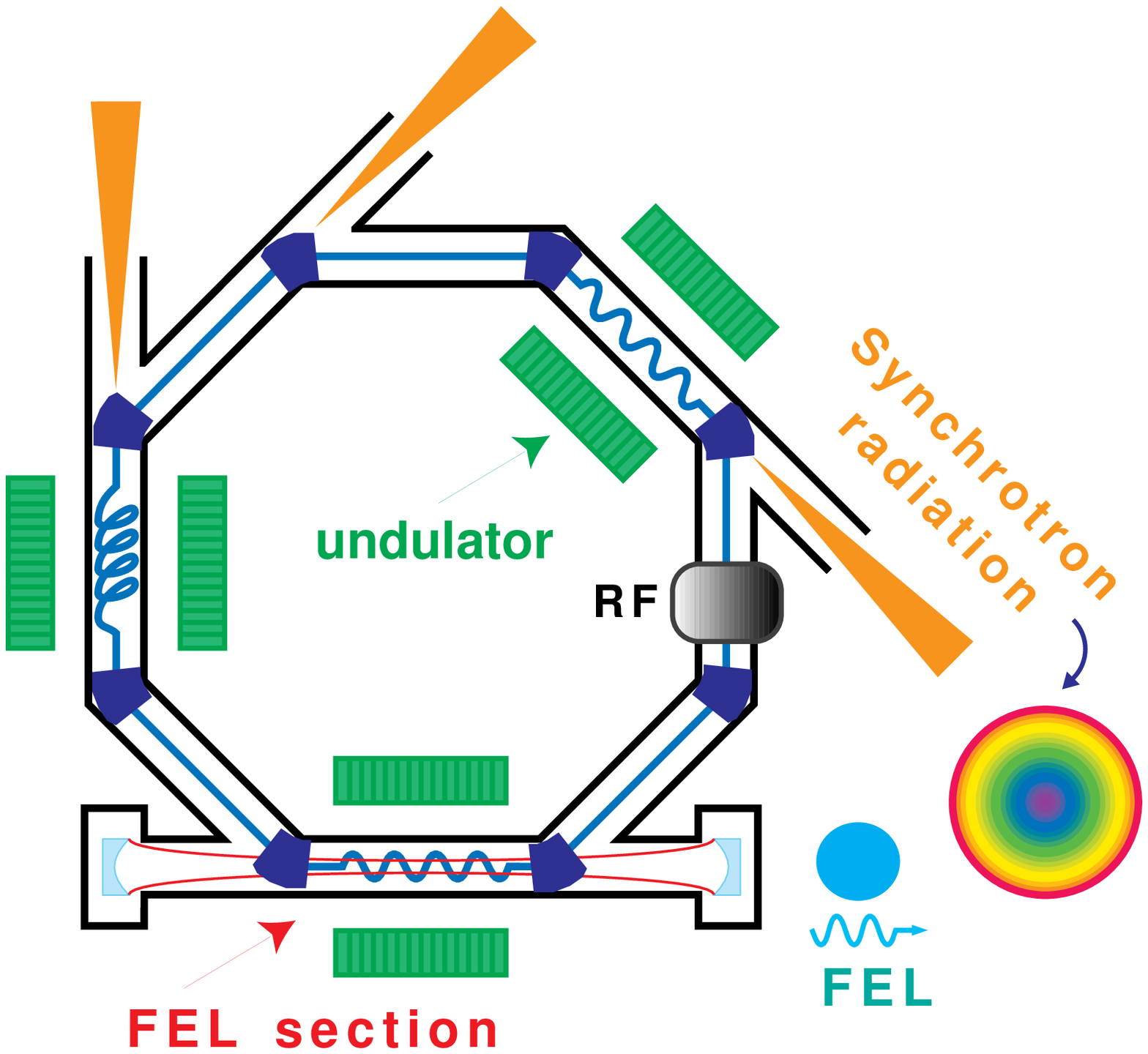, height=10truecm,width=11truecm}
\caption{\label{ane} \em Schematic layout of a SRFEL.}

\end{figure}

Since its advent, the laser has evolved as a unique tool both for 
fundamental investigations and important applications in optical science and 
technology \cite{lase}. In providing coherent and intrinsically stable emission, 
lasers result fundamentally different from conventional light sources, which are 
characterized by a signal that is composite of random and uncorrelated 
emissions.
Anyway, the view of the laser emission as an ordered and time-invariant 
process does not provide a complete description of the whole picture.
During the last twenty years, profound mathematical discoveries have 
revolutionized the understanding of nonlinear science. Lasers, in particular, 
have been found to exhibit a rich variety of nonlinear dynamical behaviors, 
including unstable or even chaotic regimes. As in the case of other 
nonlinear dynamical systems, the transition from stable to 
chaotic dynamics is obtained by varying a control parameter, such like 
the losses of the cavity \cite{are1}, its optical length \cite{glo1} 
or the gain of the amplification process \cite{bill}. The transitions are found to follow well-defined 
paths, regardless 
of the peculiar characteristics of the considered system.
These universal ``signatures'' motivated both  experimentalists and theoreticians 
in the search for physical systems  exhibiting such phenomena, and in the 
investigation of their intrinsic similarity. 
\\
Concerning lasers, first experimental studies have been carried on a 
$\mbox{CO}_2$ laser \cite{are1}, \cite{glo1}. 
It has been found that the modulation at 
a frequency $f$ of the chosen control parameter may result not only in oscillations with frequency $nf$, being $n$ 
an integer number, but also in a response at the sub-harmonics $f/n$. 
When the order $n$ of the sub-harmonics grows indefinitely, the response of 
the laser becomes irregular, though the system remains deterministic. A quite 
impressive similitude has been remarked between the ``multi-furcation'' diagram of 
the laser and that of the logistic map \cite{logi}. 
In order to reproduce these experimental results, a simple theoretical model has
been used, resting on the dynamical interplay between the electromagnetic field
within the optical cavity, and the variables of the material employed 
as ``active medium''.  Theoretical work has 
been also done for demonstrating that the chaotic behavior of a laser can be 
stabilized either by using a self-controlling feedback procedure 
\cite{fee1}, \cite{fee2} 
or by a further modulation of a control parameter \cite{fee3}, \cite{fee4} .
\\
Experimental evidences of a close link with deterministic chaos has also been 
given in the pioneering work \cite{bill}  
for a kind of non-conventional laser source: a storage-ring Free Electron Laser
(SRFEL). 
\\
After the first operation of a FEL in the infrared spectral 
range on a linear accelerator in 1977 \cite{deacon}, a second FEL was 
installed on the storage ring ACO and provided the first visible 
radiation in 1983 \cite{aco1}. 
Present Storage Ring based FELs still supply 
the shortest FEL wavelength in the oscillator configuration (on 
ELETTRA \cite{detelettra}) and in the harmonic generation scheme 
(on Super-ACO \cite{praze}). 
User applications performed since 1993 on the Super-ACO FEL have demonstrated 
the very good quality offered by such sources in terms of tunability, high 
average power, rather short pulse duration and high degree of 
coherence \cite{appl1}. 
These characteristics make, for example, the FEL very suitable for two 
colours experiments performed in combination with the naturally 
synchronized synchrotron radiation \cite{appl2}.
\\
A SRFEL, the principle of which is shown in Figure \ref{ane}, is a coherent source of 
radiation in which the active medium consists of 
an ultra-relativistic electron beam, moving in a periodic magnetic undulator. 
The alternating magnetic 
field forces the electrons to move along sine-like trajectories and, 
consequently, to emit radiation, known as spontaneous emission. 
Once stored in the optical 
cavity, the radiation is amplified to the detriment of the kinetic 
energy of the electrons. This process generally 
leads to the increase of the rms value of the particles energy 
(the so-called electron beam energy spread) and, as a consequence, 
to the reduction of the amplification gain, until this latter reaches the level 
of the cavity losses.
Since it origins from synchrotron radiation, the SRFEL output presents a 
micro-temporal structure with a period of the order of hundred of 
nanoseconds, which is determined both by the longitudinal dimension of the 
electron bunch and by the beam revolution period. 
On a larger (millisecond) temporal scale, the SRFEL dynamics depends 
strongly on the longitudinal overlap between the electron bunch(es) and the 
laser pulses at each pass inside the optical cavity (detuning condition). 
As it is shown in Figure \ref{fig2} for the particular case of Super-ACO, a 
given detuning leads to a cumulative delay between the electrons and the 
laser pulses: the laser intensity may then appear ``cw'' (for a weak or 
strong detuning) or show a stable pulsed behavior (for intermediate 
detuning) \cite{detusaco}, \cite{detuvsor}, \cite{detelettra}.
The narrow ``cw'' zone of the detuning curve (few ps around the 
perfect synchronism) is generally the most interesting for user 
applications. In fact, when in this zone, the laser is characterized 
by the maximum average power and the signal is the most close to the 
Fourier limit \cite{saco2000}. In order to keep the laser-electron 
beam synchronism and avoid the jittering, which could determine a 
migration towards one of the 
unstable, pulsed zones of the detuning curve, efficient feedback 
systems have been implemented on the Super-ACO \cite{feed1} and UVSOR 
FELs\cite{feed2}.  
SRFELs are complex, strongly coupled 
dynamical systems.
The strong laser-electron beam coupling origins 
from the fact that, unlike a LINAC based Free Electron Laser, where the beam is 
renewed after each interaction, electrons are re-circulated. As a 
consequence, at every light-electron beam energy exchange, the system 
keeps memory of previous interactions.
\\
\\
The work that will be presented in this paper starts from the results obtained 
in 1990 on the Super-ACO FEL \cite{bill}. In the latter reference, it was 
experimentally shown 
that a periodic  modulation at a given frequency $f$ of the laser-electron beam detuning 
may lead to a period doubling of the laser intensity (i.e. to a laser response at a 
frequency $f/2$) or even to chaos for more important modulation amplitudes. 
Reference \cite{bill} contains also a first attempt to define the 
conditions (i.e. the amplitude and the frequency of the  
modulation) for which the bifurcation most likely occurs. 
The experimental data were discussed in connection 
with a simplified model based on a set of phenomenological rate equations, accounting 
for the coupled evolution of the laser intensity, of the electron-beam energy spread and 
of the gain of the amplification process.  Such a model allowed to reproduce the basic 
features of the laser intensity evolution (in particular the saturation mechanism) only 
close to the perfect synchronism between the laser pulse and the electron bunch. 
This was a limiting factor for developing a comprehensive theoretical picture.
In fact, the fundamental role played by the detuning 
has been only later understood: in \cite{bill1} such effect 
has been shown to be responsible for the behavior of the laser intensity 
(``cw'' or pulsed) on the millisecond scale and to produce a further gain reduction 
over that induced by the increase of the electron-beam energy spread. 
The simple model presented in 
reference \cite{bill1} was then improved by including the detuning effect on the whole 
laser intensity distribution.   
This allowed to find a qualitative agreement between experiments and numerical 
results, for the case of the Super-ACO FEL \cite{noi}. 
\\
It is worth to stress that the loss of the laser-electron beam longitudinal 
overlapping may be induced by different phenomena such like a vibration of the cavity 
mirrors at the line frequency \cite{uvi} or a modulation of the electron 
beam energy \cite{datt1}. 
\\
In this paper an improved version of the model presented in \cite{bill1} is used to get a 
deeper insight (with respect to that obtained in 
\cite{bill}) into the physics of deterministic chaos in a SRFEL. Numerical simulations are 
compared to a set of experiments performed on the Super-ACO FEL. 
In Section 2 the model is presented and shown to be able to reproduce quantitatively the 
observed features of the detuned FEL dynamics. Section 3 discusses the modifications induced 
to the dynamics by assuming an externally modulated detuning. Bifurcations, as well
multi-furcation, are shown to occur for both the intensity and centroid 
position of the laser distribution. 
In Section 4 experimental results are presented.
Finally, Section 5 contains concluding remarks and perspectives.

\section{Theoretical model}
\label{model}
The longitudinal dynamics of a SRFEL can generally be described by a system of 
rate equations accounting for the coupled evolution of 
the electromagnetic field and of the longitudinal parameters of 
the electron bunch \cite{bill1}. 
\\
The temporal profile of the laser intensity, $y_n$, is updated at each pass, $n$,
inside the optical cavity according to:
\begin{equation}
\label{pass-pass}
y_{n+1} (\tau) = R^2 y_n (\tau-\epsilon) \left[ 1 + g_{n}(\tau)
\right] +i_s(\tau),
\end{equation}
where $\tau$ is the temporal position with respect to the 
centroid of the electron bunch distribution;
$R$ is the mirror reflectivity; the detuning parameter $\epsilon$ is the
difference between the electrons revolution period (divided by the number of 
bunches) and the period of the photons inside the cavity; $i_s$ accounts for the 
spontaneous emission of the optical klystron\footnote{ The case of SRFELs
implemented on an optical klystron is here considered \cite{vino}, \cite{elle2}. 
An optical klystron consists of two undulators separated by a dispersive section,
(i.e. a strong magnetic field) favoring the interference between the emission of
the two undulators.}.
The FEL gain $g_{n}(\tau)$ is given by:

\begin{equation}
\label{gain01}
g_{n}(\tau) =  g_i \frac{\sigma_{0}}{\sigma_{n}}\exp \left[ -
\frac{\sigma_{n}^2 - \sigma_{0}^2}{2 \sigma_{0}^2}\right]
\exp \left[ -\frac{\tau^2}{2 \sigma_{\tau,n}^2} \right]
\end{equation}
where $g_i$ and $\sigma_{0}$ are the initial (laser-off) peak gain
and beam energy spread, while 
$\sigma_{n}$ and $\sigma_{{\tau},n}$  are the 
energy spread and the bunch length after the n$th$ light-electron beam 
interaction.
The first exponential in the right-hand side of equation (\ref{gain01}) accounts 
for the modulation rate of the optical-klystron spectrum\footnote{The laser-off 
peak gain has been optimized by assuming $N+N_{d} =
1/(4\pi \sigma_{0})$ in the expression of the modulation rate, where $N$ is the 
periods' number of the undulators of the optical klystron and $N_{d}$ is the 
interference order due to its dispersive section \cite{elle1}.},
while the second one reproduces the temporal profile of the electron bunch 
distribution. The bunch distribution is therefore assumed to keep its 
``natural'' Gaussian profile under the action of the laser onset. 
This hypothesis entails that the interaction of the 
electron beam with the ring environment \cite{mic}, \cite{pot1}, \cite{pot2} 
is neglected. This important point will be further discussed with 
particular concern to the case of Super ACO.
\\
Defining $g_{n,0}$ as the peak gain after the n$th$ interaction, 
$g_{n}(\tau)$ can be written in the form:

\begin{equation}
\label{gain02}
g_{n}(\tau) = 
g_{n,0} \exp \left[-\frac{\tau^2}{2 \sigma_{\tau,n}^2} \right].
\end{equation}

Figure \ref{fig1} shows a schematic layout of the light-electron beam 
interaction in presence of a finite detuning $\epsilon$.
\\
The evolution of the normalized laser-induced energy spread
$\Sigma_{n}=(\sigma_{n}^2 - \sigma_0^2)/(\sigma_e^2 - \sigma_0^2)$ is given by:
\begin{equation}
\Sigma_{n+1}=\Sigma_{n} + \frac{2 \Delta T}{\tau_{s}}(I_{n} -
\Sigma_{n})
\end{equation}
where $\sigma_e$ is the equilibrium value (i.e. that reached at the 
laser saturation) of the energy spread at the perfect
tuning and $\Delta T$ is the bunching period of the laser 
inside the optical cavity;
$I_{n}$ is the normalized laser intensity defined as
$I_{n}=(1/I_{e})\int^{\infty}_{-\infty} y_{n}(\tau) d\tau$ (being 
$I_{e}$ the equilibrium value) and $\tau_s$
stands for the synchrotron damping time. Assuming  
that the saturation is achieved when the peak gain is equal to the cavity 
losses, $P$, the following relation holds\footnote{As function of the 
cavity losses, the mirror reflectivity $R$ is given by $\sqrt{1- P}$.}:
\begin{equation}
\label{Pgeq}
P= g_i \frac{\sigma_{0}}{\sigma_{e}}\exp \left[ -
\frac{\sigma_{e}^2 - \sigma_{0}^2}{2 \sigma_{0}^2}\right].
\end{equation}

By inserting equation (\ref{Pgeq}) in equation (\ref{gain02}) and recalling the
definition of $\Sigma_n$, a closed expression for the peak gain is 
obtained:
\begin{equation}
\label{gain}
g_{n,0} = g_i \frac{\sigma_0}{\sigma_{n}}
\left[\frac{P}{g_i}\right]^{\Sigma_{n}}
\left(\frac{\sigma_e}{\sigma_0}\right)^{\Sigma_{n}}.
\end{equation}

\begin{figure}[ht!]
\centering
\epsfig{file=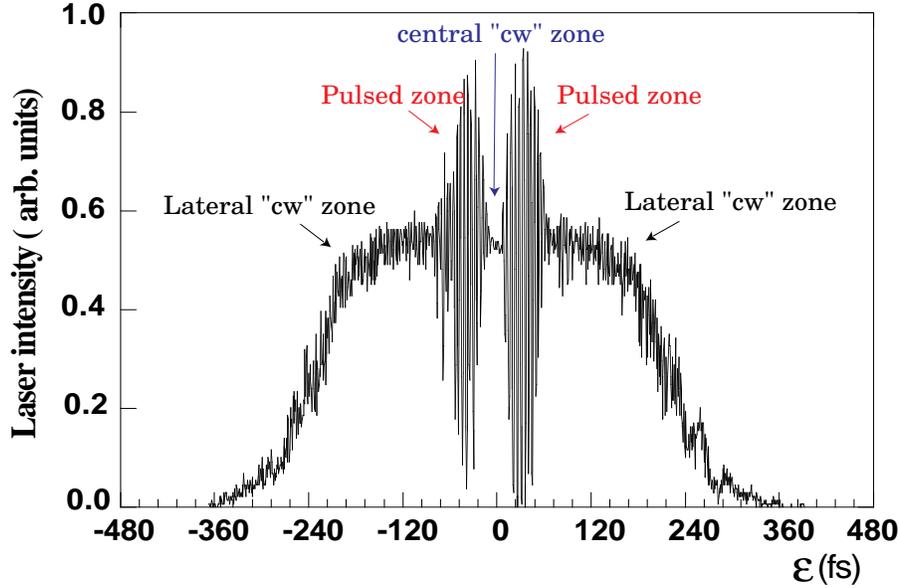,height=8truecm,width=12truecm}
\caption{\label{fig2} \em A typical detuning curve (i.e. the laser 
intensity as a function of the laser-electron beam detuning amount) obtained for the case of 
the Super-ACO FEL operated in a configuration with only the main 
RF cavity. The synchronisation between the laser pulse and the electron bunch is 
changed by means of a modification of the RF frequency (a variation of 1 Hz 
inducing a laser-electron beam detuning of 1.2 fs). The employed 
experimental conditions are those reported in Table 1, with a total 
beam current of about $40 mA$. }
\end{figure}

\begin{figure}[ht!]
\centering
\epsfig{file=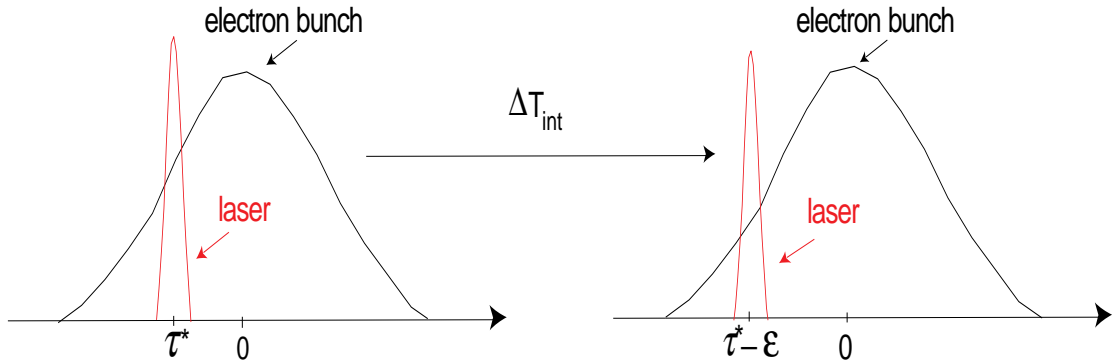, height=5truecm,width=15truecm}
\caption{\label{fig1} \em Schematic layout of the pass-to-pass 
laser-electron beam interaction. $\Delta T$ stands for the period between 
two successive interactions, $\tau^*$ is the position 
of the laser centroid with respect to the peak of the electron density 
and $\epsilon$ accounts for the laser-electron beam detuning at 
each pass.}

\end{figure}

Note that in the derivation of eq. (\ref{gain}) the 
variation of the bunch length during the FEL interaction
has been taken into account. In this respect, the present model 
represents an improvement of the one proposed in \cite{bill1}, where 
the additional assumption $\sigma_n = \sigma_0$ in the definition of 
$g_{n,0}$ is made. A systematic numerical comparison allowed to 
show that this simplification may alter the period of the laser intensity in the 
pulsed regime, that is, as it will be shown in the next Sections, a parameter of 
paramount importance for the study the laser response to an 
external modulation. 
\\
Figure \ref{fig2} shows a typical detuning curve obtained for the case of 
the Super-ACO FEL operated with only the main radio-frequency 
(RF) cavity. The parameters characterizing this configuration, which is the 
one of concern for the experiments reported in this paper, are listed in 
Table 1.

\begin{table}[h!]
\begin{center}
\vspace{5mm}
\begin{tabular}{ |c| |c| }
\hline 
\it {The Super-ACO FEL}&\\
\hline\hline
Beam energy (MeV)&800\\
\hline
Laser-off bunch length $\sigma_{\tau,0}$ (rms, ps)& 85\\
\hline
Laser-off beam energy spread $\sigma_0$ (rms) &$5.4\cdot 10^{-4}$\\
\hline
Synchrotron damping time $\tau_{s}$ (ms)&$8.5$\\
\hline
Laser width at perfect tuning (rms, ps)&20\\
\hline
Laser width at the maximum detuning (rms, ps)& 40\\
\hline
Laser wavelength (nm) &350\\
\hline
Pulse period (two bunches operation) $\Delta T$ (ns)&120\\
\hline
Laser-off peak gain $g_{i}$ (\%)&$\sim$ 2.5\\
\hline
Cavity losses $P$ (\%)&$\sim$ 0.5\\
\hline
\end{tabular}
\end{center}
\caption{\label{tab1} \em Main parameters of the Super-ACO FEL operated with only 
the main (100 MHz) RF cavity.}
\end{table}
  
\begin{figure}[ht!]
\centering
\epsfig{file=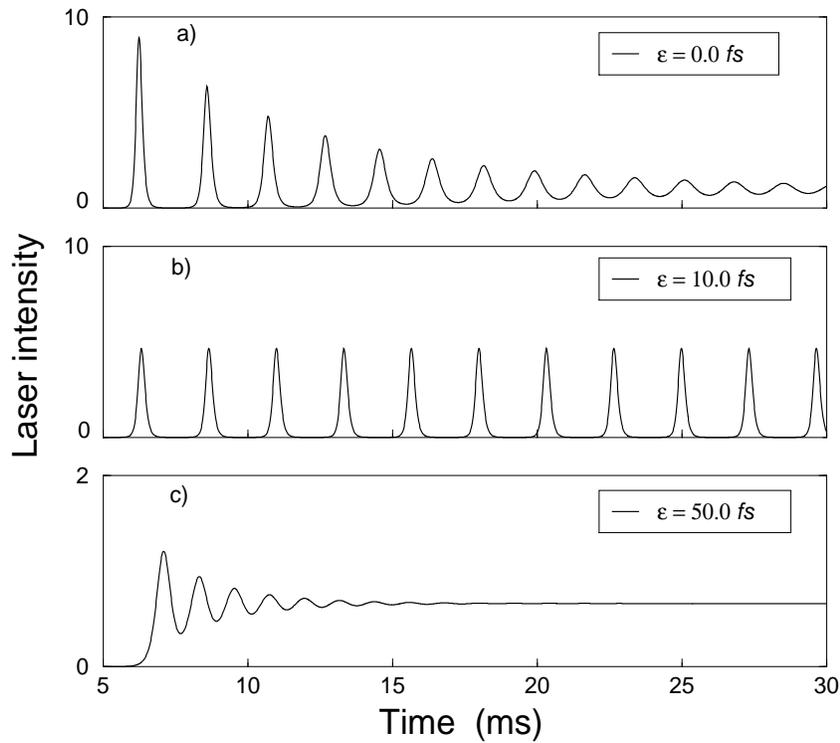,height=10truecm,width=11truecm}
\caption{\label{fig3} \em Numerical results obtained by making use of 
the parameters listed in Table 1 and reproducing the different 
``natural'' regimes of the laser intensity displayed in Figure 2. 
The ``cw'' regime in the nearly zero-detuning region (see Figure a)) 
is found to have an extension of (about) 10 fs ($\pm$ 5 fs around the 
perfect tuning). The stable pulsed regimes, observed for 
an intermediate (positive and negative) detuning amount (see Figure b)), 
has an extension of (about) 35 fs. The lateral ``cw'' regimes (observed for 
large detuning values (see Figure c)), are found to have an 
extension of the order of few hundred of fs. 
Note that these theoretical findings reproduce quantitatively the experimental results 
reported in Figure 2. The laser intensity is normalised to the equilibrium value 
it reaches in the central ``cw'' zone of the detuning curve.}    
\end{figure}

\begin{figure}[ht!]
\centering
\epsfig{file=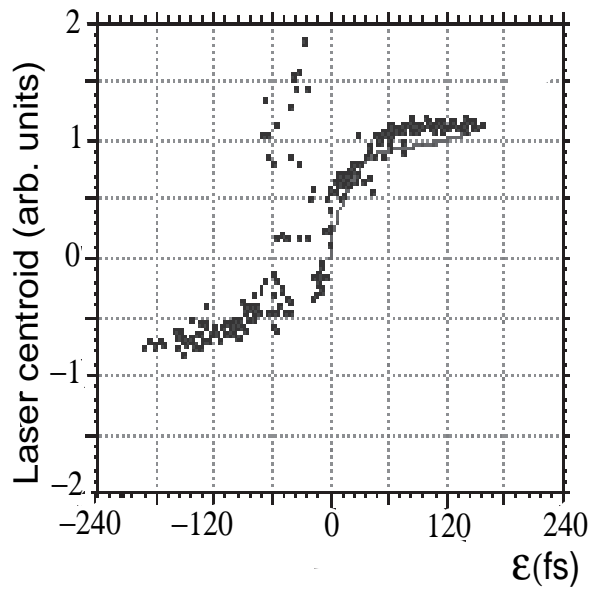,height=8truecm,width=8truecm}
\caption{\label{fig4} \em Experimental behaviour of the position of the laser 
centroid with respect to the detuning amount $\epsilon$. The employed 
experimental conditions are listed in Table 1.}
\end{figure}

\begin{figure}[ht!]
\centering
\epsfig{file=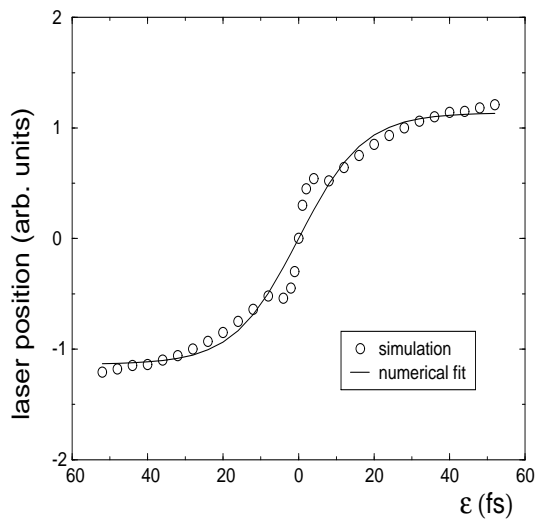,height=7truecm,width=7truecm}
\caption{\label{fig5} \em Theoretical behaviour of the position of the 
laser centroid with respect to the detuning amount $\epsilon$. Open circles refer 
to the simulations, while the solid line is obtained by making use of 
the fitting function $A \tanh (B \epsilon)$, with $A,B$
free parameters. 
}
\end{figure}

It is worth to mention that Super ACO can be also operated making use of 
an additional RF (harmonic) cavity \cite{harmo}, which has been 
installed with the aim of improving the FEL performances.
Its effect results in a shortening of the electron bunch. The consequent increase 
of the electron density leads to the enhancement of the gain of the amplification 
process. 
On the other hand, the interaction of the electron beam with the ring 
environment (i.e. the metallic 
wall of the ring vacuum chamber) is also reinforced. This leads to a 
degradation of the electron beam quality (i.e. ``anomalous'' increase of 
the laser-off bunch length vs. current and deformation of temporal beam 
distribution) 
limiting in part the 
beneficial effect of electron-density increase \cite{preraph}.
As regards the shape of the electron beam temporal distribution, 
the interaction with the ring vacuum chamber 
induces a perturbation of the ``natural'' Gaussian profile of the electron beam: 
the higher the cavity voltage, the stronger becomes the head-tail 
effect \cite{head} deforming the electron distribution. 
It is worth to stress that, due to relative high value of the Super-ACO vacuum chamber 
impedance \cite{nim}, the anomalous bunch lengthening and 
the perturbation of the beam profile are significant even when the harmonic cavity is passive.
For the purpose of this paper, these perturbations can however be 
neglected in first approximation.  
\\
A quasi-symmetric beam profile reflects in a detuning curve that is almost 
symmetric for positive and negative detuning amounts (see Figure \ref{fig2}).
The structure of the detuning curve has been studied by making use of the 
theoretical model presented above. The extension of the central (``cw'') and 
lateral (pulsed) zones, which are the ``playground'' of the experiments that will
be discussed in the next Sections, has been found to be well 
reproduced (see caption of Figure \ref{fig3}). 
Figure \ref{fig3} shows the regimes of the laser intensity 
for different 
detuning amounts. Again, the numerical results have been found in  
quantitative agreement with experiments.
Concerning, in particular, the pulsed regime, a careful experimental and numerical 
analyses \cite{noi}
pointed out that the period of the laser-intensity oscillations is not a constant 
(as it has been implicitly assumed in \cite{bill}) but it is instead an 
almost linear 
function of the detuning amount. 
\\
The effect of an imperfect light-electron beam longitudinal overlapping also 
influences the equilibrium position of the laser centroid. As shown in Figure \ref{fig4}
\cite{garz}, the laser position is quite sensitive to small detuning amounts, while it 
attains an almost constant value as the side zones of the detuning curve 
are approached.  
\\
Numerical simulations are in agreement with experiments (see Figure 
\ref{fig5}) and allow to detail the interesting behavior of the laser 
centroid in the close-to-zero detuning region.

\section{Numerical simulations for modulated detuning}

\begin{figure}[ht!]
\centering
\epsfig{file=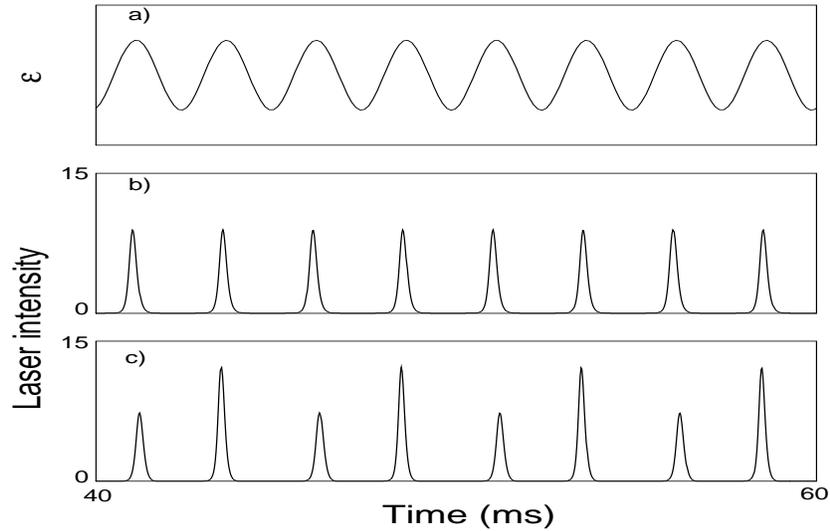, height=7truecm,width=11truecm}
\caption{\label{400Hz} \em Transition from a $1T$ to a $2T$ regime, for 
the laser 
intensity . Here $f=400~Hz$ and $b=1~fs$. Figure a) shows the modulation $\epsilon$ 
versus time. Figure b) refers to $a=19~fs$, while the Figure c) to $a=34~fs$.}
\end{figure}

\begin{figure}[ht!]
\centering
\epsfig{file=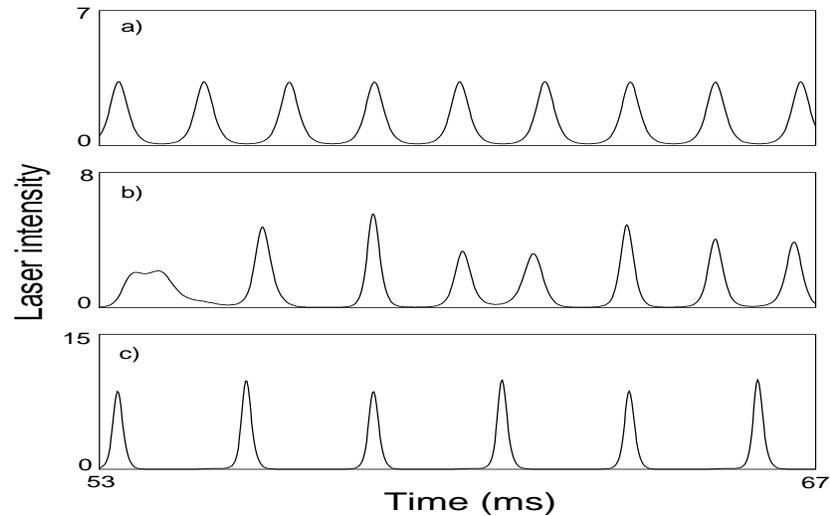, height=7truecm,width=11.5truecm}
\caption{\label{600SIM} \em Numerical simulations performed for 
$f=600~Hz$. Figure a) shows the laser intensity vs. time for $a=10~fs$ 
and $b=2~fs$. The system is locked to the frequency of the external 
modulation, thus displaying a $1T$ regime. Figure b) represents the 
laser intensity vs. time, for $a=20~fs$ and $b=2~fs$. An a-periodic 
signal is found. In figure c), for $a=30~fs$ (and $b=2~fs$), the system shows a $3T$ 
structure in qualitative agreement with experimental results of Figure
\ref{600EXP}.}
\end{figure}

\begin{figure}[ht!]
\centering
\epsfig{file=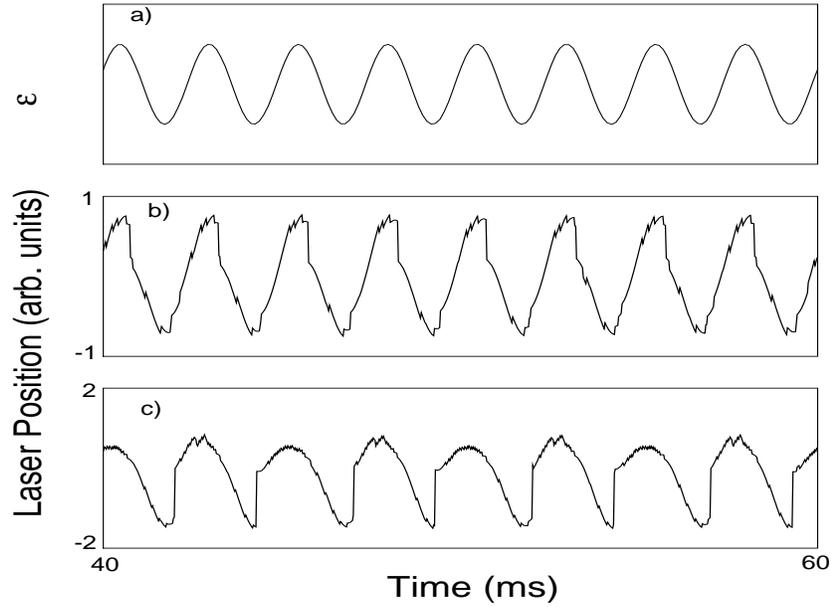, height=8truecm,width=11truecm}
\caption{\label{centroid} \em Transition from a $1T$ to a $2T$ regimes, for the position of 
the laser centroid. Here $f=400~Hz$ and $b=1~fs$, as in Figure \ref{400Hz}. 
Figure a) shows the modulation $\epsilon$ versus time. Figure b) refers 
to $a=19~fs$, while Figure c) to $a=34~fs$.}
\end{figure}

\begin{figure}[ht!]
\centering
\epsfig{file=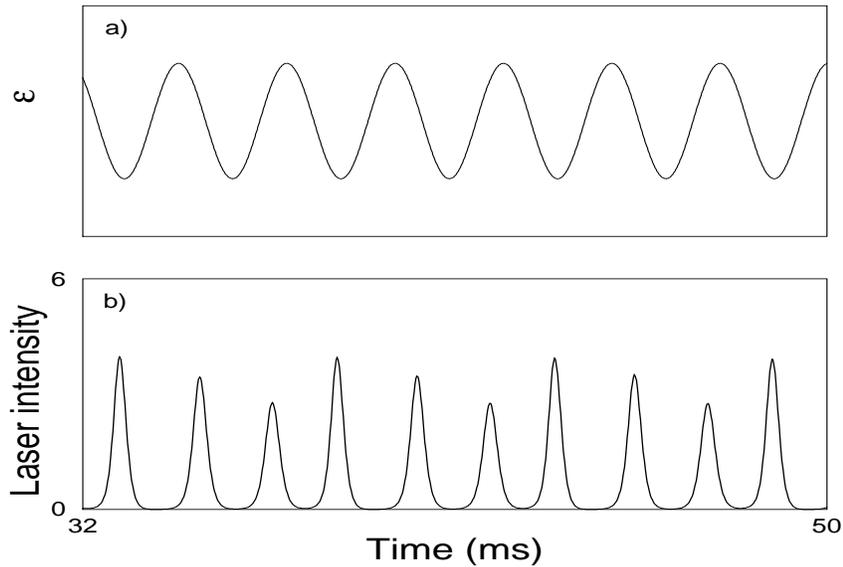, height=8truecm,width=12truecm}
\caption{\label{oro1} \em Figure a): modulation $\epsilon$ vs. time. 
Figure b): Behaviour of the laser intensity. Here, $a=2~fs$, 
$b=25~fs$ $f = 2/3 f_p = 381.6~Hz$. A $2T$ structure characterised by three peaks is displayed,
in complete agreement with the prediction of relation (\ref{oro}).  }
\end{figure}

\begin{figure}[ht!]
\centering
\epsfig{file=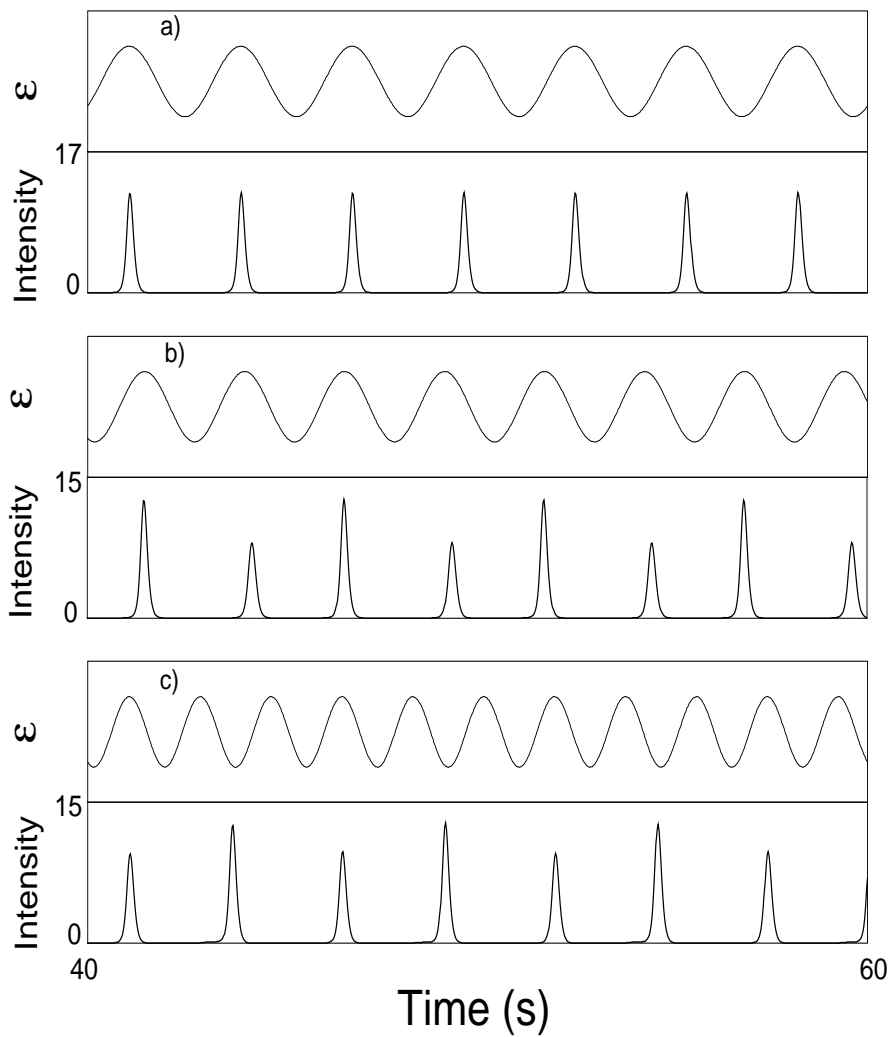, height=14truecm,width=12truecm }
\caption{\label{freq} \em Cascade $1 T  \rightarrow  2 T \rightarrow 3 T$, 
for laser intensity. Here $a=37~fs$ and $b=2~fs$. From top to the bottom, the 
frequencies are respectively $350$ $Hz$ (Figure a)), $390$ $Hz$ (Figure b)) 
and $550$ Hz (Figure c)).}
\end{figure}

Consider now the modification induced to the system dynamics by the cumulative effect of
an external periodic modulation applied to the detuning $\epsilon$, namely:

\begin{equation}
\label{modulation}  
\epsilon=a \sin(2 \pi f t) + b,
\end{equation}

where $t$ represents the time elapsed, integer multiple of  
$\Delta T$; $f$ is the frequency of the oscillation, while the two amplitudes 
$a$ and $b$ control the maximum detuning amount. The oscillations are
centered around $b$, being $a$ the maximum elongation. Tuning the value of $b$,
allows to explore the effects induced by the modulation in different zones.
Simulations have been performed based on the model previously introduced, the main 
motivation being a systematic comparison with preliminary results of experiments carried on 
in SuperACO. This point is addressed in the next Section.
This paragraph is instead devoted to a more generally-oriented analysis, concerning the 
peculiar phenomena, that have been numerically found. 
\\
\\
As already discussed in the previous Section, several 
different regimes are produced in a SRFEL, for constant detuning amounts.
For small $\epsilon$ the laser displays a ``cw'' intensity, while larger 
values of $\epsilon$ drive the system into stable oscillations. This 
``natural'' pulsed regime is of paramount importance for the present study. 
The right boundary of the central, symmetric,  ``cw'' zone will be labeled as 
$\epsilon_{th}$. 
\\
Consider first the case for which $b$ is set to zero. Hence, the average 
modulation is also zero. The external forcing, defined by equation 
(\ref{modulation}), leads to a stable ($1T$) response only for small values 
of $a$, which prevent excursion of the modulated detuning outside the 
boundaries of the ``cw'' zone. When this excursion occurs, the modulation 
is generally found to induce a chaotic response of the laser intensity.   
\\
Quite different scenario is produced when $b$ is set to some small value 
different from zero. This choice corresponds to introducing a slight 
asymmetry which, surprisingly enough, is responsible of a significant 
regularization of the signal\footnote{
Note that this case can be 
considered more realistic than the previous one. In fact, even when the 
laser is operated as close as possible to the perfect tuning, a slight, uncontrollable, 
shift 
(corresponding to a $b$ value of the order of one or two fs in the case of Super ACO) has 
to be generally expected.}.    
In Figure \ref{400Hz} the laser intensity is displayed as function of time, the 
upper panel being the modulation of $\epsilon$. Simulations refer to a fixed value 
of the frequency while the amplitude is varied. The transition from a $1T$ to a $2T$ 
clearly occurs. It is worth stressing that the latter regime is observed for $a>\epsilon_{th}$. 
\\
In Figure \ref{600SIM} the laser intensity is plotted as function of time. Again, simulations 
refer to different values of $a$, while $b$ and $f$ are maintained constant.   
For small values of $a$, the laser response is locked to the frequency of the external 
modulation, thus displaying a $1T$ regime. For larger values the 
laser intensity passes through a chaotic region and, finally, attains a stable $3T$ regime.
\\
In Figure \ref{centroid} the position of laser centroid is represented, as function of time,
for the same choice of parameters as in Figure \ref{400Hz}. Oscillations are displayed  and 
a transition from a  $1T$ to  $2T$ regimes is observed. This is, indeed, a quite general 
result: each time a bifurcation of the laser intensity takes place, an analogous response is
observed for the position of the centroid. 
\\
Summing up, a wide number of distinct behaviors are recovered and, within them, 
bifurcations of both intensity and centroid position of the laser, depending 
on the values of the parameters $a$ and $f$.
For $a<\epsilon_{th}$ , i.e. when the oscillation is confined in the 
central ``cw'' zone of the detuning curve (few ps), the system is locked to 
the frequency imposed by the 
modulation and laser intensity exhibits regular oscillations of frequency $f$.
As soon as the value of $a$ exceeds the threshold $\epsilon_{th}$, a cascade 
can occur.
Therefore, one is led to conclude that excursions in the pulsed regime are a necessary 
requirement for the modulations to produce more complex behaviors. It is, in fact, 
believed that the observed periodic structures could be the result of the
combined effect of the external and natural modulations. In addition, the system 
seems to be more sensitive to the natural frequencies (which, for the 
case of Super-ACO, lie between 400 Hz and 630 Hz), even though multi-furcations 
have been detected outside this privileged range.
These observations mark an important point of distinction with similar studies on 
conventional lasers, being the pulsed regime an intrinsic characteristic of the SRFEL.  
\\
Another interesting feature regards the value, $\hat{a}$, of the modulation amplitude 
for which the transition from $1T$ regime to more structured behaviors 
(i.e. $2T$, $3T$, \ldots) 
occurs. Numerically it has been observed that increasing the 
frequency , the value of 
$\hat{a}$ significantly decreases.
This phenomenon, 
has been systematically observed in a wide range of frequencies (i.e. 
200 Hz to 900 Hz). For larger values of $f$, the system is 
insensitive to the rapid external modulation, which therefore becomes 
ineffective.
\\
Further, numerical simulations allowed to detect  
narrow windows, for which more complex periodic structures 
(i.e. $4T$, $5T$, $6T$, $8T$...) are displayed. Nevertheless,
the existence of universal paths towards deterministic 
chaos remains an open issue. 
\\
Similar analysis have been carried out for larger values of 
$b$, such that the modulation is stably centered in one of the pulsed 
zones (see Figure \ref{oro1}). This condition is achieved for small enough values of 
the amplitude $a$.  Numerical results produce remarkable features, which 
will be commented in the following. As already pointed out in Section 2,
the frequency of the oscillation in the pulsed zone was shown to  
increase linearly with $\epsilon$ \cite{noi}. Assume $f_p$ to be
the frequency of the natural oscillation which is found for 
$\epsilon=b$, being $a=0$. When the modulation of $\epsilon$ is switched on 
and the condition:
\begin{equation} 
\label{oro}
\frac{n}{f}=\frac{m}{f_p}   ~~~~~~ n,m \in N,~~~n<m,~~~ \frac{n}{m} \in Q 
\end{equation}
is fulfilled, the laser shows oscillations of frequency $f/n$, 
characterized by $m$ peaks, spaced with period $2 \pi / f_p$ (see Figure 
\ref{oro1}, which refers to $n=2$ and $m=3$). 
Deviations are observed when condition (\ref{oro}) is no longer satisfied
(for instance, by progressively changing  the value of $b$). 
Note, however, that
such behavior occurs for discrete values of  $f$, and are
not maintained over a finite set of frequencies. Hence, its 
characterization in term of multi-furcation is, somehow, stretched. Further
experimental investigations will be addressed to validate these numerical 
findings.
\\
Finally, an extensive campaign of simulations has also been performed varying the frequency
$f$ of the modulation, while  keeping $a$ fixed. A small value has been again assigned 
to the parameter $b$, in such a way that the center of the oscillations falls in the 
region of ``cw'' laser. A typical results is reported in Figure \ref{freq}, where a cascade 
$1T~\rightarrow~2T~\rightarrow~3T$ is shown to occur.  It is again 
worth to stress the crucial role that seems to be played by the pulsed zone.

\section{Experimental results and their interpretation}

\begin{figure}[ht!]
\centering
\epsfig{file=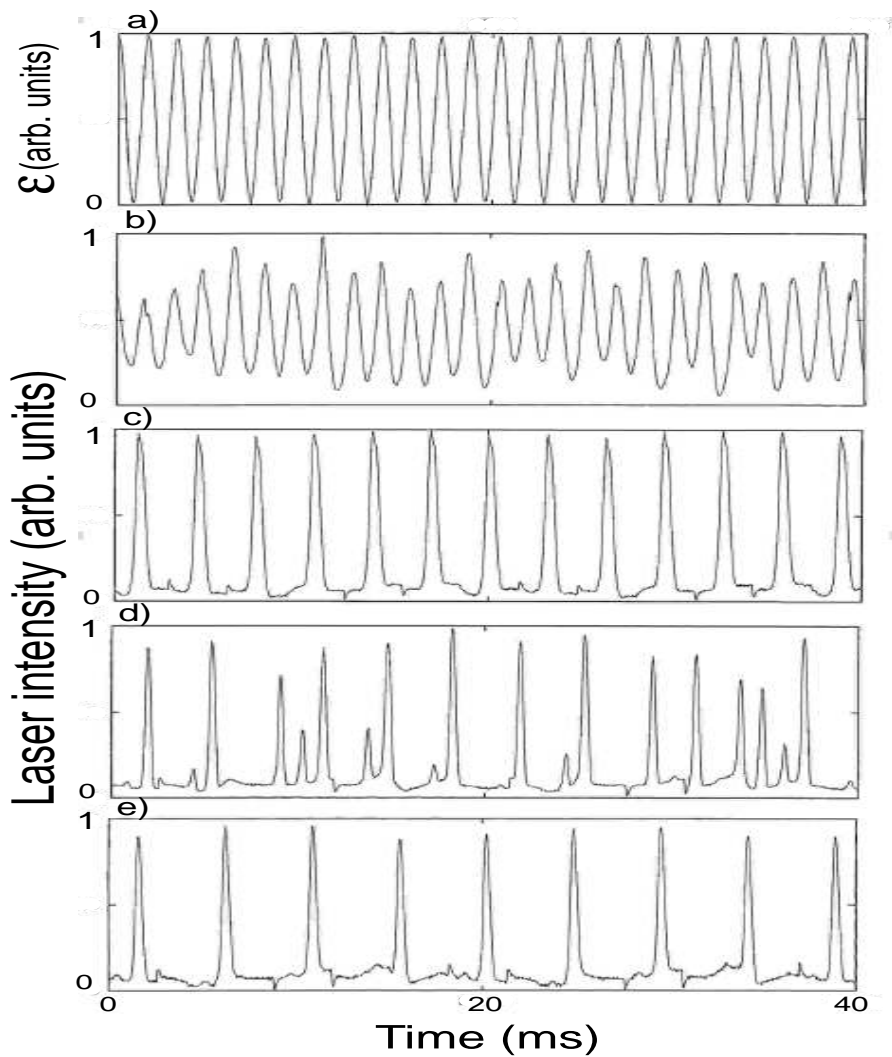, height=14truecm,width=12truecm}
\caption{\label{600EXP} \em Experimental response of the laser 
intensity to the detuning modulation. Here $f=660~Hz$. Figure a) 
shows the modulation $\epsilon$ versus time. Figure b) represents the 
laser intensity vs. time for $a=7~fs$. The system displays a 
(noisy) $1T$ regime. In figure c), for $a=12~fs$, the laser intensity 
shows a $2T$ regime. In figure d), for $a=20~fs$, an a-periodic 
signal is found. In figure e), for $a=46~fs$, the laser intensity 
attains a stable $3T$ regime.} 
\end{figure}

\begin{figure}[ht!]
\centering
\epsfig{file=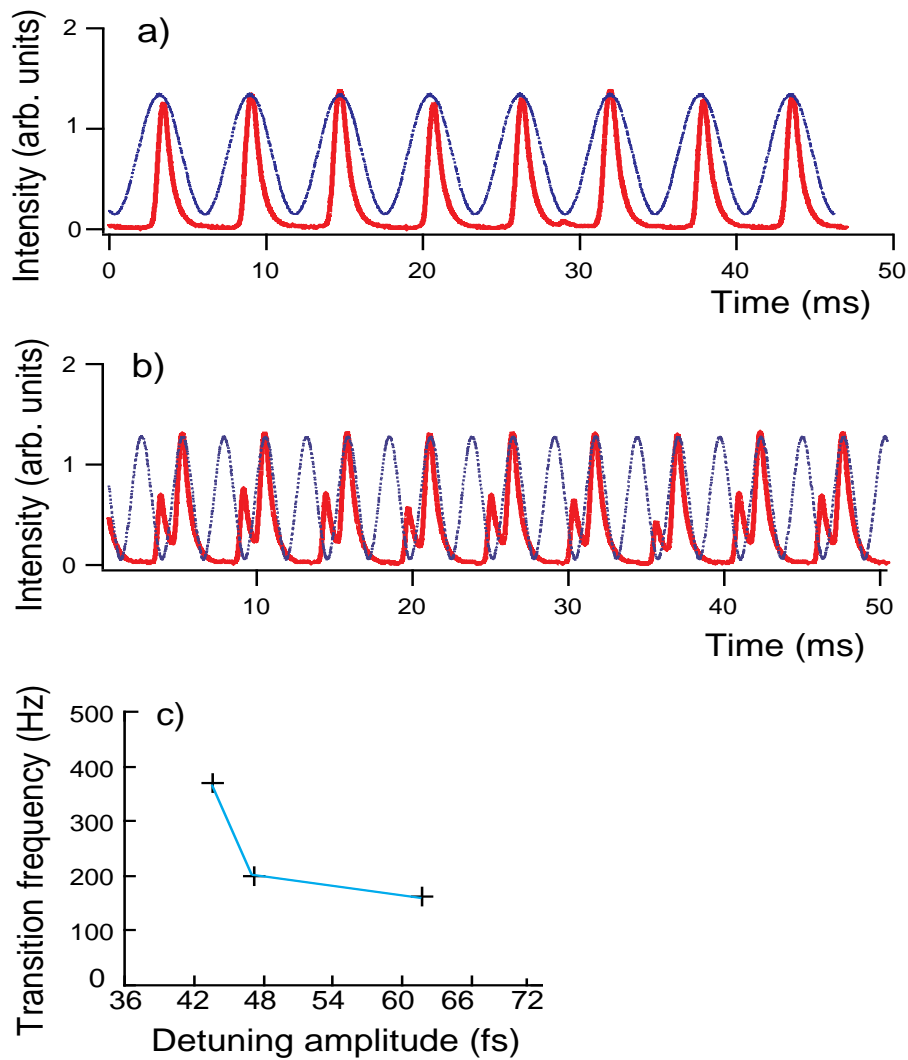, height=14truecm,width=12truecm}
\caption{\label{freqEXP} \em Experimental response of the laser 
intensity to the detuning modulation. Here the amplitude $a=42~fs$. Figure a) 
shows the laser intensity vs. time (thick solid line) for $f=174~Hz$.
The thin solid line represents the detuning modulation (arbitrary units).
The system displays a $1T$ regime. In Figure b) the laser  intensity (thick line) 
is represented vs. time, for  $f=377~Hz$. Again, the thin line refers to  
$\epsilon$. A clear $2T$ response is found to occur.
In Figure c), the values of the frequency for which the system experiences
the transitions from a $1T$ to a $2T$ regime, are plotted vs. the amplitude 
$a$. 
The decreasing tendency is clearly displayed. 
}
\end{figure}

\begin{figure}[ht!]
\centering
\epsfig{file=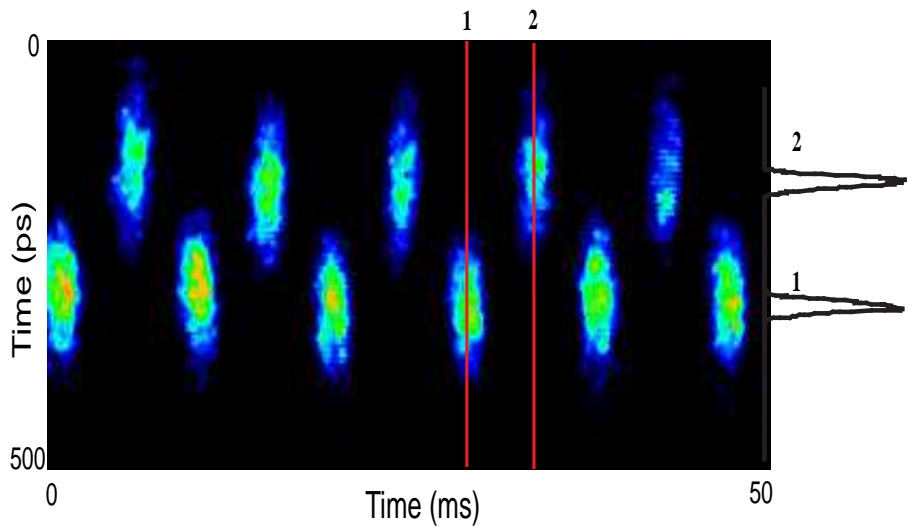, height=7truecm,width=12truecm}
\caption{\label{streak} \em Streak camera image of the Super-ACO FEL 
showing a $2T$ regime of the position of the laser centroid.  
Here $f=250~Hz$ and the modulation amplitude has been 
chosen large enough to exceed the central ``cw'' zone of the detuning 
curve. A vertical 
cut of the image provides the laser longitudinal distribution while on the horizontal 
axis one can follow the evolution in time of the distribution profile.} 
\end{figure}

Experiments have been performed by modifying the longitudinal overlap between the 
laser pulse 
and the electron bunch. This is done by modulating the 
radio-frequency in such a way that the variation of the detuning 
$\epsilon$, in the form specified by equation (\ref{modulation}), 
is achieved. 
Different results have been obtained by varying either the modulation amplitude 
(keeping $f$ constant) or the modulation frequency 
(for fixed value of the amplitude of the modulation). 
\\
In Figure \ref{600EXP} the evolution of the laser intensity is investigated, when  
adopting the first procedure: $f$ is set to a value close to the largest natural frequency. 
For small amplitudes of the modulation, a noisy $1T$ regime is observed. A transition 
toward  a $2T$ response is produced when $a$ is larger than $\epsilon_{th}$, in qualitative 
agreement with the general picture outlined in the previous Section. Larger values of $a$ 
induce a chaotic response, before a stable $3T$ regime is attained (see Figure \ref{600SIM},
for a qualitative comparison). 
\\                                  
In Figure \ref{freqEXP},  a $1T~\rightarrow~2T$ transition is observed when varying the modulation 
frequency. Again, the bifurcation has been found to occur for an amplitude of the modulation
large enough to drive the laser in the pulsed zone of the detuning curve. The transition 
is observed when $f$ approaches the lowest natural frequency. 
This result is in a good agreement  with the numerical simulations shown in 
Figure \ref{freq}, which suggest that a further bifurcation $2T~\rightarrow~3T$  
could have been found, for larger frequencies. In addition, the frequency
for which the transition is shown to occur, increases when the amplitude of the 
modulation is reduced (see Figure \ref{freqEXP}). This result confirms the prediction of the 
theoretical analysis.
\\
The response of the FEL to an external modulation of the laser-electron beam 
detuning has been also investigated by means of a double sweep streak camera. 
In particular, in Figure \ref{streak} the occurrence of a period doubling 
of the laser centroid is detected.
\\

\section{Conclusions}
\label{concl}

A complete analysis of the dynamics of a SRFEL in presence of a 
longitudinal laser-electron beam detuning has been performed. 
Numerical simulations, based on the model discussed in 
\cite{noi}, and experimental results have been compared and shown 
to agree quantitatively, for a constant value of $\epsilon$.
\\
Further, the effect of an external modulated $\epsilon$ has been 
considered. The system has been shown to display 
bifurcations, multi-furcations as well chaotic regimes for
both the laser intensity and the position of its centroid,
when either the amplitude or frequency of the 
modulation are tuned. Numerical analysis
and preliminary experiments carried on for SuperACO,
agree in this respect. Moreover, a detailed series
of simulations, over a wide range of values of the modulation
parameters, allowed to point out the crucial role played by 
the naturally pulsed zones of the SRFEL detuning curve. This observation, confirmed by 
a careful analysis of the experimental data, seems to 
indicate that the link with analogous investigations
for conventional lasers is not trivial,
the pulsed regime being an intrinsic characteristic of the 
SRFEL.
\\
In addition, conventional laser are shown to produce 
a complete cascade towards deterministic chaos, similar
to that of the logistic map. In the case of 
SRFEL, experiments proved clearly the
existence of $2T$ and $3T$ regimes, as well of
chaotic behaviors. Nevertheless, a progressive
increase of the periodicity of the structures was 
not observed. 
Numerically, the interstitial 
regions laying between  stable signals
and  chaotic regimes have been object of a detailed analysis. 
More complex periodic structures have been clearly 
detected ($4T$, $5T$, $6T$, $8T$...), even though, the 
attempt of identifying recurrent paths has, 
so far, failed. However, it is worthwhile stressing
that the latter are shown to hold for a narrow 
range of tunability of the parameters. Hence, 
the transition from a low periodicity signal to 
chaos is generally sharp. 
\\
Finally, a complete understanding of the laser
response in presence of an externally modulated detuning 
could allow to improve the performances of the 
FEL. It has in fact been proven in \cite{fee1}-\cite{fee4}
that the chaotic behavior of a conventional 
laser can be successfully stabilized. 
It is planned to investigate deeper this 
possibility, being interested in the
development of new self-controlling procedure.

\section*{Authorship}

The theoretical part of this work, described in section 2 and 3
above, was done by G.D.N. and D.F. These authors contributed equally
to this part of the work. The experimental part, described in section
4 above, was done by C.B. and M.E.C.

\section*{Acknowledgments}
\label{s:acknowledgements}
We thank M. Billardon, P. Glorieux and D. Dangoisse for stimulating 
discussions. We gratefully acknowledge the help of D. Garzella, D. Nutarelli
S. Randoux, R. Roux and B. Visentin for carrying on experiments. 
We also thank E. Aurell for a careful reading of the manuscript.  
The work of D.F. was supported by the Swedish Research Council through grant NFR F~650
-19981250. Financial support to the work of G.D.N. was provided by the
TMR contract N° ERB 4061 PL 97-0102.


\newpage

\section*{Figures}

\vspace{2truecm}

\end{document}